\documentclass[a4paper,aps,prd,10pt,preprintnumbers,showpacs,twocolumn,superscriptaddress,nofootinbib,amsmath,amssymb]{revtex4-1}
\usepackage{graphicx}
\usepackage{cmap}
\usepackage[utf8]{inputenc}
\usepackage[T1]{fontenc}

\def\K{{\cal K}}

\DeclareMathAlphabet{\pazocal}{OMS}{zplm}{m}{n}
\def\dfrac#1#2{\frac{\displaystyle #1}{\displaystyle #2}}

\begin{document}

\title{Analytic expressions for quasinormal modes of the general parametrized spherically symmetric black holes and the Hod's proposal}
\author{Alexey Dubinsky}\email{dubinsky@ukr.net}
\affiliation{Pablo de Olavide University, Seville, Spain}

\begin{abstract}
Using an expansion in terms of the inverse multipole number and the WKB approach, we derive an analytic expression for generic parametrized spherically symmetric and asymptotically flat black holes described by the Rezzolla-Zhidenko spacetime, in the regime when deviations from the Schwarzschild geometry are relatively small, i.e., when the coefficients of the parametrization are small. As an application of this analytic formula, we demonstrate that such generic black holes satisfy Hod's proposal, which constrains the damping rate of the least damped mode in relation to the Hawking temperature. 
\end{abstract}

\maketitle

\textbf{Introduction.}  The eikonal regime of quasinormal modes (QNMs) \cite{Kokkotas:1999bd,Nollert:1999ji,Konoplya:2011qq} refers to the high-frequency limit of the oscillations of black holes and other compact objects, where the wavelength of the perturbation is much smaller than the characteristic size of the system. In this regime, the behavior of QNMs can be understood using a geometric optics or "ray tracing" analogy, where the perturbations essentially follow null geodesics, or the paths that light rays would take in the curved spacetime. In the eikonal regime, the multipole number $\ell$ is large. Since $\ell$ is associated with angular momentum, large $\ell$ corresponds to perturbations with short wavelengths compared to the size of the black hole. The perturbations in this regime are akin to light rays orbiting the black hole. The quasinormal modes are related to the properties of these orbits \cite{Cardoso:2008bp,Konoplya:2017wot,Konoplya:2022gjp,Bolokhov:2023dxq}, particularly the light ring, which is the orbit where light can circle the black hole. The null geodesic is in correspondence also with the critical formation of a black hole \cite{Ianniccari:2024eza,Hod:2024ihh}.

Another intriguing aspect related to the eikonal regime is the phenomenon of eikonal instability \cite{Gleiser:2005ra,Takahashi:2011du,Konoplya:2017lhs,Takahashi:2009xh,Konoplya:2017ymp,Takahashi:2010gz}. Counterintuitively, this instability can develop at high $\ell$ in some theories of gravity, where higher multipoles may not only raise the peak of the potential but also deepen the negative gap.

Fortunately, the eikonal regime is also the limit in which quasinormal modes can be obtained analytically using the WKB method, which is accurate in this regime. Typically, quasinormal frequencies cannot be found in analytic form and require numerical computations, even for some lower-dimensional black holes \cite{Skvortsova:2023zmj,Konoplya:2020ibi}. Therefore, analytic formulas for quasinormal modes of various black holes were obtained in a great number of works and the corresponding discussions of the null geodesics/eikonal quasinormal modes, eikonal instability etc. were discussed (see, for instance \cite{Zhidenko:2008fp,Konoplya:2005sy,Konoplya:2001ji,Chen:2022nlw,Allahyari:2018cmg,Bolokhov:2023bwm,Dubinsky:2024aeu,Dubinsky:2024gwo,Konoplya:2001ji} and references therein).  
Additionally, in the eikonal regime, quasinormal modes in most cases do not depend on the spin of the field, and the system usually approaches this regime quickly, even at intermediate values of $\ell$. 

Therefore, it is not surprising that expanding beyond the eikonal limit has enabled the derivation of sufficiently accurate and compact analytic formulas \cite{Konoplya:2023moy}, which have recently been used in several works to find analytic expressions for quasinormal modes of black holes in various theories of gravity \cite{Malik:2024sxv,Malik:2024tuf,Malik:2024voy,2753764,Bolokhov:2023dxq,Malik:2023bxc}. In the present work, rather than focusing on a particular gravitational theory, we adopt an agnostic approach by considering perturbations of arbitrary spherically symmetric and asymptotically flat black holes, described by the Rezzolla-Zhidenko parametrization \cite{Rezzolla:2014mua}. We derive an analytic formula for quasinormal modes of this spacetime, assuming that the geometry deviates from the Schwarzschild limit relatively softly, that is, that the coefficients of the parametrization, measuring the deviating from the Schwarzschild limit, are relatively small.  

Furthermore, we apply the obtained analytic formula to demonstrate that the so-called Hod's proposal, which connects the damping rate of the fundamental mode with the Hawking temperature, is satisfied for the considered class of black holes.

\textbf{Formalism of black hole parametrization.}  The general and systematic way to consider deviations from the Schwarzschild geometry is suggested by the Rezzolla-Zhidenko parametrization \cite{Rezzolla:2014mua}. The metric of a spherically symmetric black hole can be expressed in a general form that involves two independent metric functions, $f(r)$ and $g(r)$:
\begin{equation}
ds^2=-f(r)dt^2+\frac{dr^2}{g(r)}+r^2 (d\theta^2+\sin^2\theta d\phi^2),\label{metric}
\end{equation}
It is common to reparameterize these functions using $B(r)$ and $N(r)$, where $f(r) = N^2(r)$ and $g(r) = \frac{N^2(r)}{B^2(r)}$. The radius of the event horizon, $r_0$, is defined by the condition $N(r_0) = 0$.

A comprehensive parametrization for spherically symmetric and asymptotically flat black holes was introduced in \cite{Rezzolla:2014mua} and later extended to accommodate axial symmetry in \cite{Konoplya:2016jvv,Younsi:2016azx}. This framework has been extensively applied and discussed in various studies \cite{Kocherlakota:2020kyu,Zhang:2024rvk,Cassing:2023bpt,Li:2021mnx,Ma:2024kbu,Bronnikov:2021liv,Shashank:2021giy,Konoplya:2021slg,Kokkotas:2017zwt,Yu:2021xen,Toshmatov:2023anz,Konoplya:2019fpy,Nampalliwar:2019iti,Ni:2016uik,Konoplya:2018arm,Zinhailo:2018ska,Paul:2023eep}, and here we provide a brief overview of its essential features.

Following the approach of \cite{Rezzolla:2014mua}, we introduce a dimensionless compact coordinate defined as
\[
x \equiv 1 - \frac{r_0}{r},
\]
where $x=0$ corresponds to the event horizon and $x=1$ corresponds to infinity. The metric function $N^2$ is then rewritten in terms of $x$ as $N^2 = x A(x)$, with $A(x) > 0$ for $0 \leq x \leq 1$. The functions $A(x)$ and $B(x)$ are expanded as follows:
\begin{eqnarray}\nonumber
A(x)&=&1-\epsilon (1-x)+(a_0-\epsilon)(1-x)^2+{\tilde A}(x)(1-x)^3\,,
\\
B(x)&=&1+b_0(1-x)+{\tilde B}(x)(1-x)^2\,.\label{ABexp}
\end{eqnarray}
Here, $\epsilon$ quantifies the deviation of the event horizon radius $r_0$ from the Schwarzschild value $2M$:
\[
\epsilon = \frac{2M - r_0}{r_0}.
\]
The coefficients $a_0$ and $b_0$ are related to post-Newtonian (PN) parameters by the expressions:
\[
a_0=\frac{1}{2}(\beta-\gamma)(1+\epsilon)^2,
\qquad
b_0=\frac{1}{2}(\gamma-1)(1+\epsilon).
\]
Current observational limits place $a_0$ and $b_0$ at approximately $10^{-4}$, making them typically negligible.

The functions ${\tilde A}(x)$ and ${\tilde B}(x)$, which determine the near-horizon geometry (i.e., $x \simeq 0$), are expressed as infinite continued fractions:
\begin{equation}\label{ABdef}
{\tilde A}(x)=\dfrac{a_1}{1+\dfrac{a_2x}{1+\dfrac{a_3x}{1+\ldots}}}, \quad
{\tilde B}(x)=\dfrac{b_1}{1+\dfrac{b_2x}{1+\dfrac{b_3x}{1+\ldots}}}.
\end{equation}
Here, $a_1, a_2, \ldots$ and $b_1, b_2, \ldots$ are dimensionless constants that serve as the coefficients of the parametrization. When $a_0 = b_0 = 0$, the continued fractions reduce at the event horizon to:
\[
{\tilde A}(0) = a_1, \quad {\tilde B}(0) = b_1.
\]
This implies that, near the event horizon, the higher-order terms in the expansions can often be neglected, as the leading-order terms dominate. Consequently, truncating the infinite continued fractions after a few terms typically provides a good approximation for a wide range of black hole metrics \cite{Konoplya:2020hyk}.
When all the coefficients of the parametrization: $\epsilon$, $a_i$ and $b_i$ vanish, we reproduce the Schwarzschild metric.  

\textbf{Wave equations.}  The behavior of minimally coupled scalar ($\Phi$) and electromagnetic ($A_\mu$) fields in a curved spacetime is governed by the general covariant Klein-Gordon and Maxwell equations, respectively. These equations take the following forms:
\begin{subequations}
\begin{eqnarray}\label{KGg}
\frac{1}{\sqrt{-g}}\partial_\mu \left(\sqrt{-g}g^{\mu \nu}\partial_\nu\Phi\right) &=& 0,
\\\label{EmagEq}
\frac{1}{\sqrt{-g}}\partial_{\mu} \left(F_{\rho\sigma}g^{\rho \nu}g^{\sigma \mu}\sqrt{-g}\right) &=& 0\,,
\end{eqnarray}
\end{subequations}
where $F_{\mu\nu}=\partial_\mu A_\nu-\partial_\nu A_\mu$ represents the electromagnetic field tensor, capturing the interaction of the electromagnetic field with the curved background.

By expanding the fields into spherical harmonics, the above equations, \eqref{KGg} and \eqref{EmagEq}, can be reduced to a more tractable form—a universal radial wave-like equation (see, for example, Eqs. 7 and 8 in \cite{Konoplya:2006rv}):
\begin{equation}\label{wave-equation}
\dfrac{d^2 \Psi}{dr_*^2} + \left(\omega^2 - V(r)\right)\Psi = 0,
\end{equation}
where $r_*$ is the so-called "tortoise coordinate," defined by:
\begin{equation}\label{tortoise}
dr_* \equiv \frac{dr}{\sqrt{f(r) g(r)}}.
\end{equation}
The tortoise coordinate effectively stretches the radial coordinate near the event horizon, simplifying the analysis of wave propagation in this region.

The effective potentials corresponding to scalar ($s=0$) and electromagnetic ($s=1$) fields are given by:
\begin{equation}\label{potentialScalar}
V(r) = f(r)\frac{\ell(\ell+1)}{r^2} + \frac{1-s}{r}\cdot\frac{d^2 r}{dr_*^2},
\end{equation}
where $\ell$ is the multipole number, which determines the angular structure of the perturbation, and also generates the azimuthal quantum number $m$.

For gravitational perturbations, however, the situation is more complex. In general, if we consider a parametrized black hole that is not tied to a specific theory of gravity, it becomes challenging to express the wave equations for gravitational perturbations in a universal form. Nonetheless, various theories adopt an effective approach, assuming that the dynamics of the gravitational field can be described within the framework of Einstein's theory of gravity, but with an effective energy-momentum tensor that represents an anisotropic fluid \cite{Ashtekar:2018lag,Ashtekar:2018cay}.

Such an approach is justified because there exist modified theories of gravity, sometimes referred to as "Einsteinian theories," where black hole solutions arise from anisotropic fluids \cite{Dehnen:2003rc,Ivashchuk:2010sn}. In these theories, axial gravitational perturbations (corresponding to $s=2$) can be simplified into a wave equation similar to the scalar and electromagnetic cases. The effective potential for these axial perturbations takes the form (see, for instance, \cite{Bouhmadi-Lopez:2020oia,Konoplya:2024lch}):
\begin{equation}
V(r) = f(r)\left(\frac{2g(r)}{r^2} - \frac{(fg)'}{2rf} + \frac{(\ell+2)(\ell-1)}{r^2}\right).
\end{equation}

\begin{figure*}
\resizebox{0.9 \linewidth}{!}{\includegraphics{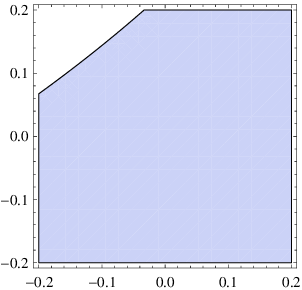}~\includegraphics{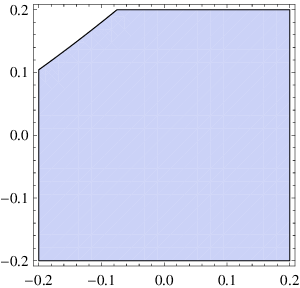}~\includegraphics{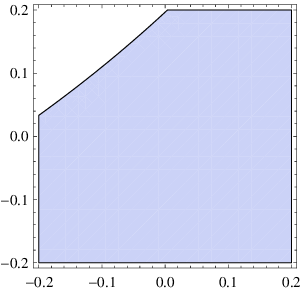}}
\caption{The shaded part of the plot shows the region where $\pi T_{H} - |Im (\omega)|$ is positive definite, From left to right: $b_{1}=0$, $b_{1}=-0.1$, $b_{1}=0.1$; abscissa stands for $a_1$ and ordinate is for $\epsilon$.}\label{fig:Hod}
\end{figure*}

Interestingly, despite the different physical origins of the fields (scalar, electromagnetic, and gravitational), the effective potentials for these perturbations can be unified into a single, universal form:
\begin{align}\nonumber
V_{s}(r) = & \frac{(1-s) \left(g(r) f'(r) + f(r) g'(r)\right)}{2r} - \\
& \frac{(1-s) s (f(r) g(r) - f(r))}{r^2} + \frac{\ell (\ell + 1) f(r)}{r^2}.
\end{align}
This expression shows how the potentials depend on the spin $s$ of the field: $s=0$ for scalar, $s=1$ for electromagnetic, and $s=2$ for gravitational perturbations. The first term in this equation captures the interaction between the background geometry (through $f(r)$ and $g(r)$) and the field, while the second and third terms contribute spin-dependent corrections and angular momentum barriers, respectively. This unified treatment underscores the versatility of the radial wave equation in capturing the essential dynamics of different fields in curved spacetimes.

\begin{table}
\centering
\begin{tabular}{ccc}
 \hline
 $\epsilon$ & WKB & analytic \\
 \hline
 \multicolumn{3}{c}{$a_{1}=b_{1}=0$} \\ 
 \hline
 -0.1 & 1.01906-0.219889 i & 1.02007-0.217397 i \\
 -0.05 & 0.99343-0.206452 i & 0.993654-0.204924 i \\
 0. & 0.967286-0.193518 i & 0.96724-0.19245 i \\
 0.05 & 0.940666-0.181203 i & 0.940826-0.179976 i \\
 0.1 & 0.913628-0.169625 i & 0.914412-0.167503 i \\
 \hline
 \multicolumn{3}{c}{$a_{1}=b_{1}=0.1$} \\ 
 \hline
  -0.1 & 1.03323-0.217131 i & 1.03467-0.213925 i \\
 -0.05 & 1.00763-0.203943 i & 1.0082-0.201509 i \\
 0. & 0.981511-0.191172 i & 0.981727-0.189092 i \\
 0.05 & 0.954876-0.178904 i & 0.955254-0.176676 i \\
 0.1 & 0.927767-0.167251 i & 0.92878-0.164259 i\\
 \hline
  \multicolumn{3}{c}{$a_{1}=b_{1}=-0.1$} \\ 
 \hline
  -0.1 & 1.00489-0.223038 i & 1.00568-0.219691 i \\
 -0.05 & 0.979274-0.209332 i & 0.979248-0.207242 i \\
 0. & 0.953151-0.196263 i & 0.952815-0.194794 i \\
 0.05 & 0.926594-0.183953 i & 0.926382-0.182346 i \\
 0.1 & 0.899715-0.172517 i & 0.899948-0.169898 i\\
\hline
  \multicolumn{3}{c}{$a_{1}=0.1, \quad b_{1}=-0.1$} \\ 
 \hline
  -0.1 & 1.03211-0.239129 i & 1.03402-0.235315 i \\
 -0.05 & 1.00702-0.224268 i & 1.0079-0.221264 i \\
 0. & 0.981427-0.209837 i & 0.981792-0.207213 i \\
 0.05 & 0.955237-0.195863 i & 0.95568-0.193161 i \\
 0.1 & 0.928433-0.182546 i & 0.929568-0.17911 i\\
\hline
\end{tabular}
\caption{Fundamental quasinormal modes ($n=0$) for scalar field perturbations at $\ell=2$ calculated by the 6th order WKB method \cite{Konoplya:2003ii} with Padé approximants \cite{Matyjasek:2017psv} and via the obtained analytic formula (\ref{analytic}).}
\end{table}
\begin{table}
\centering
\begin{tabular}{ccc}
 \hline
 $\epsilon$ & WKB & analytic \\
 \hline
 \multicolumn{3}{c}{$a_{1}=b_{1}=0$} \\ 
 \hline
 -0.1 & 0.960379-0.215402 i & 0.962855-0.217397 i \\
 -0.05 & 0.938131-0.202508 i & 0.939387-0.204924 i \\
 0. & 0.91519-0.190009 i & 0.91592-0.19245 i \\
 0.05 & 0.89158-0.178018 i & 0.892452-0.179976 i \\
 0.1 & 0.867345-0.166657 i & 0.868985-0.167503 i\\
 \hline
 \multicolumn{3}{c}{$a_{1}=b_{1}=0.1$} \\ 
 \hline
 -0.1 & 0.976373-0.213043 i & 0.979201-0.213925 i \\
 -0.05 & 0.954018-0.200394 i & 0.955615-0.201509 i \\
 0. & 0.930987-0.188061 i & 0.932029-0.189092 i \\
 0.05 & 0.90728-0.176129 i & 0.908443-0.176676 i \\
 0.1 & 0.882915-0.164713 i & 0.884857-0.164259 i\\
 \hline
  \multicolumn{3}{c}{$a_{1}=b_{1}=-0.1$} \\ 
 \hline
 -0.1 & 0.944359-0.218238 i & 0.946959-0.219691 i \\
 -0.05 & 0.922254-0.205069 i & 0.923517-0.207242 i \\
 0. & 0.899451-0.192416 i & 0.900075-0.194794 i \\
 0.05 & 0.875994-0.180402 i & 0.876634-0.182346 i \\
 0.1 & 0.851969-0.16916 i & 0.853192-0.169898 i\\
\hline
  \multicolumn{3}{c}{$a_{1}=0.1, \quad b_{1}=-0.1$} \\ 
 \hline
 -0.1 & 0.968532-0.23444 i & 0.972991-0.235315 i \\
 -0.05 & 0.947325-0.220261 i & 0.95016-0.221264 i \\
 0. & 0.925346-0.206377 i & 0.927329-0.207213 i \\
 0.05 & 0.902571-0.192884 i & 0.904498-0.193161 i \\
 0.1 & 0.879-0.179917 i & 0.881667-0.17911 i\\
\hline
\end{tabular}
\caption{Fundamental quasinormal modes ($n=0$) for electromagnetic field perturbations at $\ell=2$ calculated by the 6th order WKB method \cite{Konoplya:2003ii} with Padé approximants \cite{Matyjasek:2017psv} and via the obtained analytic formula (\ref{analytic}).}
\end{table}

\textbf{WKB method.} The Wentzel-Kramers-Brillouin (WKB) method, a semi-classical approximation technique, is effectively utilized for solving wave equations with potential barriers. It relies on the asymptotic matching of WKB solutions with the Taylor expansion around the potential's peak. Specifically, the first-order WKB approximation aligns with the eikonal limit, becoming exact as the angular momentum number $\ell$ approaches infinity. The comprehensive expression for quasinormal frequencies, as elaborated within the WKB framework, is presented as an expansion around this eikonal limit:
\begin{eqnarray}\label{WKBformula-spherical}\nonumber
\omega^2 &=& V_0 + \Sigma_{i=1}^{i=\infty} A_{2 i}(\K^2) +\\
&-& i\K\sqrt{-2V_2}\left(1  + \Sigma_{i=1}^{i=\infty} A_{2 i+1}(\K^2)\right),
\end{eqnarray}
where $V_0$ is the peak of the potential barrier and $V_2$ is related to its second derivative. The coefficients $A_i$ are detailed across various references, covering up to the 13th order in the WKB approximation \cite{Iyer:1986np,Konoplya:2003ii,Matyjasek:2017psv}.

Adopting the methodology introduced in \cite{Konoplya:2023moy}, we expand the effective potential in terms of the multipole number $\ell$:
\begin{equation}\label{potential-multipole}
V(r_*) =  \kappa^2 \Sigma_{i=0}^{i=\infty} H_{i}(r_*) \kappa^{-i},
\end{equation}
where $\kappa \equiv \ell + \frac{1}{2}$, thus $\kappa$ large implies high angular momentum. The expansion terms $H_i(r_*)$ denote the hierarchy of corrections decreasing with increasing $\kappa^{-1}$.

The eikonal limit for $\K$, derived from the first-order WKB approximation, is given by the following relation:
\begin{equation}\label{eikonalK}
-i\K_0 \sqrt{-2V_2} = \omega^2 - V_0,
\end{equation}
indicating the relation under the close proximity of classical turning points:
\begin{equation}\label{deltaw}
\delta \equiv \omega^2 - \Omega\kappa^2 \approx 0.
\end{equation}
This proximity condition suggests a regime where $\omega$, the frequency of the perturbation, nearly matches the natural frequency scale set by $\Omega$ and $\kappa$. In such scenarios, $\K \approx \K_0 \approx 0$, which significantly influences the transmission coefficients, making them distinguishable from trivial limits.

Expanding $\K$ around its eikonal value $\K_0$, it is deduced that \cite{Konoplya:2023moy}:
\[
\frac{\omega^2}{\kappa^2} - \Omega^2 \equiv \frac{\delta}{\kappa^2} \approx -i\K_0\frac{\sqrt{-2V_2}}{\kappa^2} = \mathcal{O}(\kappa^{-1}),
\]
facilitates a series expansion for $\omega$, thereby refining the estimate of $\K$. This refined approach, underpinning the expansion around the eikonal approximation, provides a more accurate depiction of wave dynamics near potential barriers in various fields.

\begin{table}
\centering
\begin{tabular}{ccc}
 \hline
 $\epsilon$ & WKB & analytic \\
 \hline
 \multicolumn{3}{c}{$a_{1}=b_{1}=0$} \\ 
 \hline
 -0.1 & 1.2616-0.210209 i & 1.26816-0.209846 i \\
 -0.05 & 1.23056-0.197588 i & 1.23612-0.197934 i \\
 0. & 1.19889-0.185406 i & 1.20409-0.186021 i \\
 0.05 & 1.16661-0.173763 i & 1.17205-0.174109 i \\
 0.1 & 1.13381-0.162774 i & 1.14001-0.162197 i\\
 \hline
 \multicolumn{3}{c}{$a_{1}=b_{1}=0.1$} \\ 
 \hline
 -0.1 & 1.27923-0.207279 i & 1.28618-0.206366 i \\
 -0.05 & 1.24827-0.194945 i & 1.25411-0.194592 i \\
 0. & 1.21664-0.182951 i & 1.22204-0.182819 i \\
 0.05 & 1.18436-0.171382 i & 1.18997-0.171045 i \\
 0.1 & 1.15147-0.160344 i & 1.1579-0.159272 i\\
 \hline
  \multicolumn{3}{c}{$a_{1}=b_{1}=-0.1$} \\ 
 \hline
 -0.1 & 1.24445-0.213853 i & 1.2503-0.212021 i \\
 -0.05 & 1.21336-0.200932 i & 1.21824-0.200054 i \\
 0. & 1.18167-0.188562 i & 1.18618-0.188087 i \\
 0.05 & 1.14946-0.176856 i & 1.15412-0.176119 i \\
 0.1 & 1.11684-0.165928 i & 1.12206-0.164152 i\\
\hline
  \multicolumn{3}{c}{$a_{1}=0.1, \quad b_{1}=-0.1$} \\ 
 \hline
 -0.1 & 1.27887-0.229657 i & 1.28571-0.228007 i \\
 -0.05 & 1.24829-0.215528 i & 1.2539-0.214572 i \\
 0. & 1.21698-0.201779 i & 1.22208-0.201137 i \\
 0.05 & 1.18495-0.188506 i & 1.19027-0.187702 i \\
 0.1 & 1.15225-0.175827 i & 1.15846-0.174267 i\\
\hline
\end{tabular}
\caption{Fundamental quasinormal modes ($n=0$) for gravitational perturbations at $\ell=3$ calculated by the 6th order WKB method \cite{Konoplya:2003ii} with Padé approximants \cite{Matyjasek:2017psv} and via the analytic formula at the second order beyond eikonal approximation.}
\end{table}

This expression is quite accurate when dealing with the fundamental mode $n=0$ and $\ell > s$, where $s$ is the spin of the field, as demonstrated in Tables I-II. To verify the accuracy, we employed the 6th-order WKB method with Padé approximants \cite{Matyjasek:2017psv}. Although it shares the same WKB origin as the derived analytic formula, this method is known to be highly precise in this regime, exhibiting a relative error of a small fraction of one percent (see for instance \cite{Konoplya:2022hbl,Bolokhov:2024ixe,Zinhailo:2024kbq,Konoplya:2020jgt,Zinhailo:2024jzt,Guo:2022hjp,Churilova:2021tgn,Skvortsova:2024atk,Dubinsky:2024hmn,Gonzalez:2022ote}) when compared with the precise Leaver method \cite{Leaver:1985ax} or time-domain integration \cite{Gundlach:1993tp}. Indeed, for $\ell=2$ scalar and electromagnetic perturbations, the relative difference between the frequencies obtained using the analytic formula and those calculated with the 6th-order WKB method with Padé approximants is typically much less than one percent. However, for gravitational perturbations, the relative error is significantly larger, suggesting that an extension of the above analytic formula to one or more orders beyond the eikonal limit might be necessary (see, for example, table III for gravitational perturbations). While such an extended formula remains relatively compact, we will not present it explicitly here, as it is not essential for the further application of the analytic formula in relation to Hod's proposal. The \textit{Mathematica} code for generating expansions at various orders beyond the eikonal approximation is available from the author upon request. It is worth mentioning that the influence of higher order terms $a_2$, $a_3$, $b_2$, $b_3$, etc. on quasinormal modes is strongly suppressed as was shown in \cite{Konoplya:2022tvv}.

\textbf{Quasinormal modes.}  We can represent the position of the maximum of the effective potential in the form of the expansion in terms of powers of $\kappa^{-1}$, $\epsilon$, $a_1$ and $b_1$ and then use it the higher order WKB formula. This way we find that the position of the maximum is located at
\begin{widetext}
\begin{align}\nonumber
& r_{max} = a_{1} \left(\epsilon  \left(-\frac{4 \left(13 s^2+18 s-31\right)}{729 \kappa ^2}-\frac{4}{9}\right)-\frac{2
   \left(s^2-2 s+1\right)}{27 \kappa
   ^2}-\frac{2}{9}\right)+\epsilon  \left(\frac{5 s^2-24
   s+19}{54 \kappa ^2}+\frac{7}{18}\right)+\frac{s^2-1}{6
   \kappa ^2}+\frac{3}{2}\\
& + b_{1} \left(a_{1} \left(\frac{832 (s-1)
   \epsilon }{6561 \kappa ^2}-\frac{16 (s-1)}{243 \kappa
   ^2}\right)+\frac{92 (s-1) \epsilon }{243 \kappa
   ^2}-\frac{4 (s-1)}{9 \kappa ^2}\right)+O(\kappa^{-3}, \epsilon^2, a_{1}^2, b_{1}^2)
\end{align}
Using the above expression for $r_{max}$ in the WKB formula we find the quasinormal frequency 
\begin{align}\nonumber\label{analytic}
& \omega=-\frac{144 s^2+60 \K^2-29}{648 \sqrt{3} \kappa }+\frac{2
   \kappa }{3 \sqrt{3}}-\frac{2 i \K}{3 \sqrt{3}} +\epsilon  \left(\frac{1008 s^2+3456 s+4740 \K^2-707}{17496
   \sqrt{3} \kappa }-\frac{10 \kappa }{27
   \sqrt{3}}+\frac{70 i \K}{81 \sqrt{3}}\right)+\\\nonumber
& a_{1} \left(\epsilon  \left(\frac{-2880 s^2+9936
   s+40740 \K^2+4105}{59049 \sqrt{3} \kappa }-\frac{16
   \kappa }{729 \sqrt{3}}+\frac{128 i \K}{243
   \sqrt{3}}\right)-\frac{144 s^2+60 \K^2-65}{4374 \sqrt{3}
   \kappa }+\frac{8 \kappa }{81 \sqrt{3}}-\frac{16 i \K}{81
   \sqrt{3}}\right)+\\\nonumber
& b_{1} \left(\epsilon  \left(\frac{2 \left(540 s^2-1080
   s-1920 \K^2+79\right)}{6561 \sqrt{3} \kappa }-\frac{392 i
   \K}{729 \sqrt{3}}\right)+\frac{2 \left(-36 s^2+72 s+48
   \K^2-13\right)}{729 \sqrt{3} \kappa }+\frac{8 i \K}{27
   \sqrt{3}}\right)+\\\nonumber
& a_{1} b_{1} \left(\epsilon  \left(\frac{32
   \left(324 s^2-648 s-5748 \K^2-457\right)}{177147 \sqrt{3}
   \kappa }-\frac{5632 i \K}{19683 \sqrt{3}}\right)-\frac{32
   \left(9 s^2-18 s-39 \K^2+7\right)}{6561 \sqrt{3} \kappa
   }+\frac{128 i \K}{729 \sqrt{3}}\right) + \\
&   O(\kappa^{-2}, a_{1}^{2}, b_{1}^{2}, \epsilon^{2})
\end{align}
\end{widetext}

\textbf{Application to the Hod's conjecture.} S. Hod conjectured that the imaginary part of the quasinormal mode frequency for the least damped mode is constrained by the Hawking temperature $T_H$ of the black hole. Specifically, he proposed that:
\[
|\text{Im}(\omega)| \leq \frac{\pi T_H}{\hbar}
\]
where $\omega$ is the quasinormal mode frequency, and $\hbar$ is the reduced Planck constant \cite{Hod:2006jw,Hod:2007tb,Hod:2008zz}. This inequality suggests that the damping rate of the least damped quasinormal mode is bounded from below by a multiple of the Hawking temperature. While this conjecture has encountered counterexamples in some channels of gravitational perturbations in theories with higher curvature corrections \cite{Cuyubamba:2016cug}, other channels satisfy Hod's bound. In a real perturbation process, it is impossible to perturb only one of the channels without affecting the others. Thus, in a certain sense, we could say that Hod's conjecture has not been disproved. The Hod's bound could also be used for proving the Strong Censorship bound for quasinormal modes \cite{Hod:2020ktb}.

Here, we aim to determine whether Hod's bound is satisfied for generic spherically symmetric black holes, assuming the deviation from Schwarzschild geometry is not large, which would imply that some underlying Einstein theory has acquired a perturbative correction. It is evident that if the deviation is sufficiently small, the bound will hold, as it is satisfied in the Schwarzschild limit. The deviations we consider are not infinitesimal deformations that would render the problem trivial. Indeed, as shown in Tables I-III, the fundamental quasinormal modes deviate by more than $10\%$ from their Schwarzschild limit. Therefore, we need to establish whether Hod's bound is guaranteed for spherically symmetric and asymptotically flat black holes within some range of parameters around the Schwarzschild limit. For this purpose, we will use the obtained analytic expression for quasinormal modes of general parametrized spherically symmetric black holes.

The Hawking temperature for general spherically symmetric parametrized black hole depends on the lowest order coefficients only,
\begin{eqnarray}\label{Hod}
    T_H=\frac{A(0)}{4\pi B(0)}=\frac{1-2\epsilon+a_0+a_1}{4\pi r_0(1+b_0+b_1)}.
\end{eqnarray}
The spectrum of frequencies contains all multipole numbers $\ell$ and usually it is sufficient to check the inequality  \ref{Hod} in the eikonal regime in order to guarantee that there is at least a single mode satisfying the Hod' bound. The simplest way is to take the first order (eikonal) expansion of the generic black hole metric and extract the imaginary part of the frequency which does not depend on the spin of the field in this regime.
\begin{widetext}
\begin{equation}
\text{Im}(\omega) = -i \frac{\epsilon  (64 a_{1} (81-44 b_{1})-5292
   b_{1}+8505)+27 (8 a_{1} (8 b_{1}-9)+27 (4
   b_{1}-9))}{19683 \sqrt{3}}.
\end{equation}
Then, the inequality $\pi T_H - |\text{Im}(\omega)| \geq 0$ implies
\begin{align}\label{inequality}
& \frac{a_{1}-2 \epsilon +1}{4 (b_{1}+1)}- \frac{|27
   (27 (4 b_{1}-9)+8 a_{1} (8 b_{1}-9))+(64
   a_{1} (81-44 b_{1})-5292 b_{1}+8505)
   \epsilon |}{19683 \sqrt{3}} \geq 0
\end{align}
\end{widetext}
The  Hawking temperature must be positive, i.e.
\begin{equation}
\frac{a_{1}-2 \epsilon +1}{4 (b_{1}+1)} \geq 0.
\end{equation}
Analysis of the inequality \ref{inequality} show that in the range of relatively small values of $\epsilon$, $a_1$ and $b_1$ the Hod's bound is fulfilled, which is illustrated in fig. \ref{fig:Hod}.

\textbf{Conclusions.}
In this work, we have derived an analytic formula for the quasinormal frequencies of the general parametrized Rezzolla-Zhidenko spacetime, which provides a general description of spherically symmetric and asymptotically flat black holes. The simplest formula, developed at the first order beyond the eikonal limit, allows for the approximation of quasinormal modes even at intermediate values of $\ell$. As an application of this analytic expression, we have demonstrated that Hod's bound on the least damped quasinormal mode, in relation to the Hawking temperature, holds as long as the deviation from the Schwarzschild limit is not large. This formula can be easily extended to higher orders to achieve better accuracy. 

The analytic formula for quasinormal modes can similarly be extended to higher orders beyond the eikonal approximation and to higher-order expansions in terms of the parametrization coefficients. However, although the expressions become more complex, we find that the effects of higher-order expansions are evidently subdominant.

\textbf{Acknowledgments.}
 The author thanks R. A. Konoplya for useful discussions. The author acknowledges the Pablo de Olavide University for their support through the University-Refuge Action Plan.
\bibliography{Bibliography}
\end{document}